\documentclass[lettersize,journal]{IEEEtran}
\usepackage[final]{graphicx}
\usepackage{subfigure}
\usepackage{float}
\usepackage{amsmath}
\usepackage{cases}
\usepackage{color}
\usepackage{colortbl}
\usepackage{algorithm}
\usepackage{algpseudocode}
\usepackage{amsmath}
\usepackage{amssymb}
\usepackage{booktabs}
\usepackage{setspace}
\usepackage{array}
\usepackage{rotating}
\usepackage[normalem]{ulem}
\usepackage{bbding}

\renewcommand{\normalsize}{\fontsize{10}{12}\selectfont}

\makeatletter
\renewcommand{\maketag@@@}[1]{\hbox{\m@th\normalsize\normalfont#1}}%
\makeatother
\usepackage{stfloats}
\usepackage{cite}
\usepackage{makecell}
\usepackage{multirow}
\usepackage{hyperref}

\ifCLASSINFOpdf
\else
\fi

\usepackage{caption}
\usepackage{subcaption}
\usepackage{mathtools}
\usepackage{etoolbox}  

\begin{document}
\title{Beyond Target-Level: ISAC-Enabled Event-Level Sensing for Behavioral Intention Prediction}
\author{Haotian Liu,~\IEEEmembership{Student Member,~IEEE,}
Zhiqing Wei,~\IEEEmembership{Member,~IEEE,}
Yucong Du,~\IEEEmembership{Student Member,~IEEE,}
Jiachen Wei,~\IEEEmembership{Student Member,~IEEE,}
Xingwang Li,~\IEEEmembership{Senior Member,~IEEE,}
Zhiyong Feng,~\IEEEmembership{Senior Member,~IEEE}

\thanks{Haotian Liu, Zhiqing Wei, Yucong Du, Jiachen Wei, and Zhiyong Feng are with the Key Laboratory of Universal Wireless Communication, Ministry of Education, Beijing University of Posts and Telecommunications, Beijing 100876, China (emails: \{haotian\_liu; weizhiqing; duyc; weijiachen; fengzy\}@bupt.edu.cn). \textit{Corresponding authors: Zhiqing Wei, Zhiyong Feng.}

Xingwang Li is with the School of Physics and Electronic Information Engineering, Henan Polytechnic University, Jiaozuo 454000, China (e-mail: lixingwangbupt@gmail.com).}}

\maketitle

\begin{abstract}
Integrated Sensing and Communication (ISAC) holds great promise for enabling event-level sensing, such as behavioral intention prediction (BIP) in autonomous driving, particularly under non-line-of-sight (NLoS) or adverse weather conditions where conventional sensors degrade. 
However, as a key instance of event-level sensing, ISAC-based BIP remains unexplored. To address this gap, we propose an ISAC-enabled BIP framework and validate its feasibility and effectiveness through extensive simulations. 
Our framework achieves robust performance in safety-critical scenarios, improving the F1-score by 11.4\% over sensor-based baselines in adverse weather, thereby demonstrating ISAC's potential for intelligent event-level sensing. 
\end{abstract}

\section{Introduction}
Integrated Sensing and Communication (ISAC) is a key 6G technology that enables reliable communication and ubiquitous sensing by reusing communication infrastructure~\cite{Deng2025}. 
Ubiquitous sensing aims to understand the environment at three levels: 
target-level sensing for detecting and tracking physical states (e.g., in autonomous driving, vehicle velocity and position); event-level sensing for interpreting spatiotemporal interactions to recognize or predict behaviors (e.g., lane change or braking intention), a level that provides semantic support for advanced decision-making compared to target-level sensing; and situation-level sensing for performing global reasoning over multiple entities for holistic scene understanding (e.g., traffic flow analysis), which enables system-wide optimization and resource management.
Current ISAC research focuses on target-level sensing, achieving progress in high-precision localization, tracking, and imaging~\cite{liu2025cooperative,Liu2022,Liu2024target}. However, this has not yet fully exploited ISAC’s potential as a networked, intelligent sensing paradigm.

Extending ISAC to event-level sensing is poised to bring transformative advantages to applications such as autonomous driving and smart factories. 
Leveraging ISAC cooperative sensing, which provides wide-area coverage, multi-view observation, and all-weather operation in the sub-6 GHz band~\cite{liu2025cooperative,Liu2024target}, event-level sensing can accurately identify or predict interactive behaviors under challenging conditions, such as non-line-of-sight (NLoS) \textcolor{blue}{and} adverse weather, where conventional sensors degrade~\cite{Benrachou2022}.  
Standard vehicle-to-network (V2N) communication, while enabling interactive behavior sensing, fundamentally relies on the same onboard sensors and lacks the infrastructure-based, resilient sensing capability of ISAC to overcome these NLoS/adverse weather limitations~\cite{cheng2022}.
Nevertheless, the shift from target-level sensing to event-level sensing introduces key modeling complexities: efficiently extracting spatiotemporal interaction features, capturing long-term temporal dependencies, and fusing heterogeneous, asynchronous multi-source data.

Research in autonomous driving offers useful insights. 
Behavioral intention prediction (BIP) studies employ recurrent neural networks, bidirectional long short-term memory (Bi-LSTMs), and Transformers to model complex spatiotemporal interactions among entities~\cite{Scheel2018BiLSTM,Liang2025}. 
However, these methods rely on LiDAR, millimeter-wave radar, or cameras, which suffer from line-of-sight dependency, environmental sensitivity, and limited sensing range~\cite{Benrachou2022,Fang2024BIP}. 
Moreover, while ISAC’s target-level sensing capability continues to advance, its exploration of higher cognitive levels remains in its infancy. 
For instance, some works~\cite{guo2025multiges,wan2024prospective} address gesture or posture recognition. 
However, they only classify isolated individual behaviors and fail to capture inter-agent interactions, which constitute the core of event-level sensing. 
Thus, ISAC event-level sensing remains an open problem.

This letter presents the first step toward ISAC event-level sensing. 
Using interactive BIP as a case study, we propose an ISAC-based BIP framework to investigate ISAC's feasibility for high-level sensing. Recognizing asynchronous feature fusion as a key technical bottleneck, we propose \textcolor{blue}{a} novel asynchronous spatiotemporal interaction BIP (ASI-BIP) network. Unlike standard multimodal fusion, ASI-BIP employs independent Bi-LSTMs to process multi-rate data, thereby preserving temporal fidelity without interpolation or alignment.
Extensive simulations demonstrate that our framework maintains robust prediction performance in safety-critical scenarios such as NLoS and adverse weather. 
Notably, it improves the F1-score by 11.4\% over sensor-based baselines in adverse weather, providing early evidence of ISAC’s potential for intelligent, high-level event-level sensing.

The rest of this letter is organized as follows. Section \ref{se2} introduces the ISAC system model. Section \ref{se3} proposes our ISAC-based BIP framework. Section \ref{se4} validates our framework's feasibility and effectiveness in event-level sensing through extensive simulations. This letter concludes with Section \ref{se5}. 

\textit{Notations:} $\{\cdot\}$ stands for a set. 
For example, $\{A_i\}_{i=1}^I$ denotes the set $\{A_1,A_2,\cdots,A_I\}$.
Vectors and matrices are written in bold letters and in capital bold letters, respectively. 
$\mathbb{C}$ denotes the set of complex number and 
$\left[\cdot\right]^{\text{T}}$ stands for transpose operator.

\section{System Model}\label{se2}
As shown in Fig.~\ref{fig1}, we consider a BIP scenario with a single target vehicle (TV) and an unmanned vehicle (UV), where the TV’s physical information is acquired via ISAC cooperative sensing and the UV’s state (including position, velocity, and acceleration) is provided by onboard sensors such as the Global Navigation Satellite System. Multiple ISAC base stations (ISAC-BSs) are synchronized via optical fibers~\cite{liu2025cooperative} and communicate with UV through V2N links~\cite{Fang2024BIP}.
The extension to multi-TV scenarios introduces complexities such as trajectory overlap and identity resolution, which will be explored in future work.
The task process is as follows: 
The UV requests sensing support via V2N. ISAC-BSs perform cooperative sensing to acquire the TV’s data. The network then fuses this with UV's V2N data over a $K$-second window for event-level sensing and instruction transmission.

\subsection{ISAC Sensing Signal Model}
For the $i$-th ISAC-BS ($i = 1, 2, \ldots, I$) located \textcolor{blue}{at} $(x_\mathrm{bs}^i, y_\mathrm{bs}^i)$ and equipped with $N_\mathrm{R}$ antennas, $N_\mathrm{c}$ subcarriers, and $M$ downlink OFDM symbols are used for sensing. 
Assuming a sensing refresh rate of $R_\mathrm{h}$~Hz, the BS performs $S = R_\mathrm{h} K$ sensing snapshots within the $K$-second observation window.

In the $s$-th sensing snapshot ($s = 1, 2, \ldots, S$), we assume that the true position and velocity of the TV are given by $\mathbf{v}_\mathrm{tv}=\left[x_\mathrm{tv}^s, y_\mathrm{tv}^s,v_\mathrm{tv}^{\mathrm{x},s}, v_\mathrm{tv}^{\mathrm{y},s}\right]^\mathrm{T}$. The ISAC echo signal received by the $i$-th ISAC-BS on the $n$-th subcarrier and $m$-th OFDM symbol is expressed as~\cite{Liu2024target}
\begin{equation}\label{eq1}
    \mathbf{y}_{n,m}^{i,s} = 
 b_i^se^{j2\pi f_{\mathrm{d},i}^smT}e^{-j2\pi n \Delta f \tau_i^s} \mathbf{a}_\mathrm{R}\left(\theta_i^s\right)
 + \mathbf{z}_{n,m}^{i,s},
\end{equation}
where $\mathbf{y}_{n,m}^{i,s} \in\mathbb{C}^{N_\mathrm{R}\times 1}$ and $b_i^s$ \textcolor{blue}{denote} the sum of attenuation and transmit beamforming gain; $f_{\mathrm{d},i}^s = \frac{-2v_{s,i}^\mathrm{r}f_\mathrm{c}}{c_0}$ denotes the Doppler shift, with $v_{s,i}^\mathrm{r}=v_\mathrm{tv}^{\mathrm{x},s}\cos{(\theta_i^s)}+v_\mathrm{tv}^{\mathrm{y},s}\sin{(\theta_i^s)}$, $f_\mathrm{c}$, and $c_0$ being the radial velocity, carrier frequency, and speed of light, respectively; $\Delta f$ represents subcarrier spacing, and $\tau_i^s=\frac{2R_i^s}{c_0}$ denotes the delay, with $R_i^s$ being the distance between the $i$-th BS and TV; 
$T$ denotes the total duration of an OFDM symbol;
$\theta_i^s$ is the angle of arrival (AoA), {$\mathbf{z}_{n,m}^{i,s}\sim\mathcal{CN}\left(0,\sigma_i^2\mathbf{I}\right)$ is the aggregate thermal noise and residual inter-BS interference, which is assumed to be \textcolor{blue}{independent and identically distributed} zero-mean complex Gaussian; $\mathbf{a}_\mathrm{R}(\cdot)$ is a receive steering vector, denoted by~\cite{liu2025cooperative}
\begin{equation}\label{eq2}
\fontsize{8}{8}
  \mathbf{a}_\mathrm{R}(\cdot) = \left[1,\cdots,e^{j2\pi\frac{d_\mathrm{r}}{\lambda}p\sin{(\cdot)}},\cdots,e^{j2\pi\frac{d_\mathrm{r}}{\lambda}(N_\mathrm{R}-1)\sin{(\cdot)}}\right]^\mathrm{T},  
\end{equation}
where $p\in\{0,1,\cdots,N_\mathrm{R}-1\}$, $d_\mathrm{r}$, and $\lambda$ are the \textcolor{blue}{indices} of antennas, antenna spacing, and wavelength, respectively.

\begin{figure}
    \centering
    \includegraphics[width=0.35\textwidth]{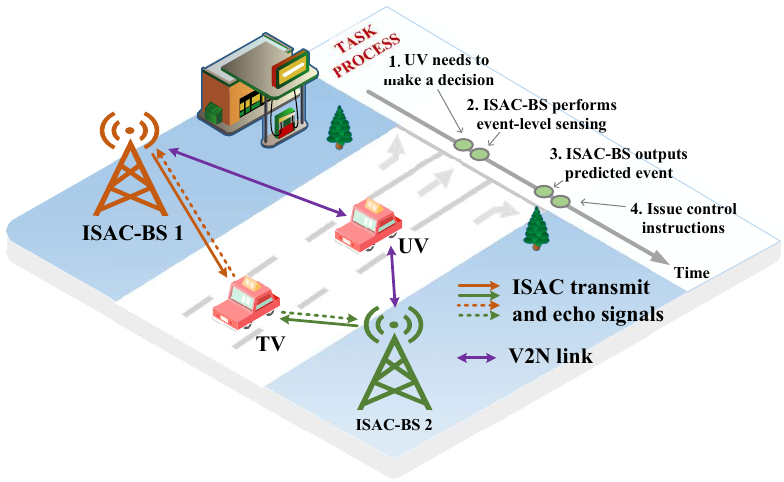}
    \caption{\underline{\textbf{(Modified)}} The illustration for the BIP based on two ISAC-BSs}
    \label{fig1}
\end{figure}

\section{Proposed framework}\label{se3}
We propose an ISAC-based BIP framework consisting of two stages: ISAC echo signal processing and BIP inference modeling.

\subsection{BIP Problem Formulation}\label{se3-a}
We formulate the BIP problem as a temporal classification task, where the TV's behavioral intentions can include: ``\textit{hard brake}'', ``\textit{left lane change}'', ``\textit{right lane change}'', ``\textit{overtake}'', ``\textit{hard accel}'', ``\textit{evasive swerve}'', and ``\textit{following}''. 
Our goal is to predict the TV's behavioral intention based on the physical information matrices of both the TV and the UV within the $K$-second observation window.

Within the $K$-second observation window, the TV's physical information matrix obtained via ISAC cooperative sensing is expressed as 
\begin{equation}
    \mathbf{P}_\mathrm{tv} = \left[\mathbf{x}_\mathrm{tv}(1), \cdots, \mathbf{x}_\mathrm{tv}(s), \cdots, \mathbf{x}_\mathrm{tv}(S)\right]\in\mathbb{C}^{6\times S}
\end{equation}
where $\mathbf{x}_\mathrm{tv}(s) = \left[\hat{x}_\mathrm{tv}^{s},\hat{y}_\mathrm{tv}^s,\hat{v}_\mathrm{tv}^{\mathrm{x},s},\hat{v}_\mathrm{tv}^{\mathrm{y},s},\hat{a}_\mathrm{tv}^{\mathrm{x},s},\hat{a}_\mathrm{tv}^{\mathrm{y},s}\right]^\mathrm{T}$ denotes the estimated TV's position, velocity, and acceleration at time $s/R_\mathrm{h}$.
Note that $\{\hat{a}_\mathrm{tv}^{\mathrm{x},s},\hat{a}_\mathrm{tv}^{\mathrm{y},s}\}$ are derived from $\{\hat{x}_\mathrm{tv}^{s},\hat{y}_\mathrm{tv}^s,\hat{v}_\mathrm{tv}^{\mathrm{x},s},\hat{v}_\mathrm{tv}^{\mathrm{y},s}\}$ using a six-dimensional Kalman filter with a constant acceleration (CA) model (\textcolor{blue}{detailed} in Section \ref{se3-b}).
The UV collects its own physical information at a lower refresh rate $R_\mathrm{l}$, yielding $C=R_\mathrm{l}K$ sensing snapshots in the observation window and the corresponding physical information matrix is 
\begin{equation}
   \mathbf{P}_\mathrm{uv}=\left[\mathbf{x}_\mathrm{uv}(1), \cdots, \mathbf{x}_\mathrm{uv}(c), \cdots, \mathbf{x}_\mathrm{uv}(C)\right]\in\mathbb{C}^{6\times C} 
\end{equation}
where $\mathbf{x}_\mathrm{uv}(c)$ defined analogously.
Since $R_\mathrm{h}>R_\mathrm{l}$, the TV and UV matrices are sampled at different rates and \textcolor{blue}{are} thus inherently asynchronous.

This asynchrony complicates \textcolor{blue}{the} direct fusion or correlation of the two matrices, making it challenging to capture their dynamic interactions.
Therefore, the performance of BIP depends critically on both the accuracy of the  physical information matrices and the fidelity of interaction modeling between \textcolor{blue}{them}.

\subsection{ISAC Echo Signal Processing}\label{se3-b}
This section aims to provide \textcolor{blue}{the} high-precision physical information matrix of the TV for BIP. 
As an illustrative example, the $s$-th sensing processing involves the following steps.
While the individual steps build upon existing ISAC techniques, their systematic combination and specific adaptation to generate the complete, high-precision, six-dimensional state vector ($\mathbf{x}_\mathrm{tv}(s)$) required for event-level sensing is novel and serves as a prerequisite for our subsequent inference model.

\textbf{Step 1:} For the $i$-th ISAC-BS, the received echo signal on $N_\mathrm{c}$ subcarriers and $M$ OFDM symbols is organized into a tensor $\mathcal{Y}_{i,s}\in\mathbb{C}^{N_\mathrm{R}\times N_\mathrm{c}\times M}$. 
The grid-based three-dimensional discrete Fourier transform method proposed in~\cite{Liu2024target} is then applied to estimate the angle $\hat{\theta}_i^s$, delay $\hat{\tau}_i^s$, and Doppler shift $\hat{f}_{\mathrm{d},i}^s$.
After processing all $I$ ISAC-BSs, we obtain the corresponding sets of estimates: $\{\hat{\theta}_i^s\}_{i=1}^I$, $\{\hat{\tau}_i^s\}_{i=1}^I$, and $\{\hat{f}_{\mathrm{d},i}^s\}_{i=1}^I$.

\textbf{Step 2:} The positioning problem is formulated as a maximum likelihood estimation (MLE) problem using hybrid time-difference-of-arrival (TDoA)/AoA method, denoted by
\begin{equation}\label{eq3}
\resizebox{0.40\textwidth}{!}{$ %
\begin{gathered}
-
\begin{bmatrix}
2x_{2,1} & 2y_{2,1} & 2r_{2,1} \\
2x_{3,1} & 2y_{3,1} & 2r_{3,1} \\
\vdots & \vdots & \vdots \\
2x_{I,1} & 2y_{I,1} & 2r_{I,1} \\
-\sin\left(\hat{\theta}_1^s\right) & \cos\left(\hat{\theta}_1^s\right) & 0 \\
-\sin\left(\hat{\theta}_2^s\right) & \cos\left(\hat{\theta}_2^s\right) & 0 \\
\vdots & \vdots & \vdots \\
-\sin\left(\hat{\theta}_I^s\right) & \cos\left(\hat{\theta}_I^s\right) & 0
\end{bmatrix}
\begin{bmatrix}
\textcolor{blue}{x_\mathrm{tv}^s} \\
\textcolor{blue}{y_\mathrm{tv}^s} \\
\frac{\hat{\tau}_1^sc_0}{2}
\end{bmatrix}=
\begin{bmatrix}
r_{2,1}^2-k_2+k_1 \\
r_{3,1}^2-k_3+k_1 \\
\vdots \\
r_{I,1}^2-k_I+k_1 \\
\sin\left(\hat{\theta}_1^s\right)x_{\mathrm{bs}}^1-\cos\left(\hat{\theta}_1^s\right)y_{\mathrm{bs}}^1 \\
\sin\left(\hat{\theta}_2^s\right)x_{\mathrm{bs}}^2-\cos\left(\hat{\theta}_2^s\right)y_{\mathrm{bs}}^2 \\
\vdots \\
\sin\left(\hat{\theta}_I^s\right)x_{\mathrm{bs}}^I-\cos\left(\hat{\theta}_I^s\right)y_{\mathrm{bs}}^I
\end{bmatrix}
\end{gathered}
$}%
\end{equation}
where $x_{i,1} = x_\mathrm{bs}^i-x_\mathrm{bs}^1$, $y_{i,1} = y_\mathrm{bs}^i-y_\mathrm{bs}^1$, $k_i=(x_\mathrm{bs}^i)^2+(y_\mathrm{bs}^i)^2$, and $r_{i,1}=\frac{c_0}{2}(\hat{\tau}_i^s-\hat{\tau}_1^s)$. Eq.~\eqref{eq3} is solved via the weighted least squares algorithm. 
The weighting matrix $\mathbf{W}$ is defined as the noise covariance matrix $\mathbf{R}^{-1}$, where $\mathbf{R}=\text{diag}\left(\hat{\sigma}_2^2,\hat{\sigma}_3^2,\cdots,\hat{\sigma}_I^2,\hat{\sigma}_1^2,\hat{\sigma}_2^2,\cdots,\hat{\sigma}_I^2\right)\in\mathbb{R}^{(2I-1)\times(2I-1)}$ and $\hat{\sigma}_i^2$ can be obtained by MLE. 
\textcolor{blue}{For simplicity, we assume identical noise variances for range and angle measurements at a given BS.}

\textbf{Step 3:} Similar to Step 2, the velocity estimation is also formulated as an MLE problem by exploiting the geometric relationship between the TV's velocity vector (magnitude and orientation) and the observed Doppler shifts $\{\hat{f}_{\mathrm{d},i}^s\}_{i=1}^I$~\cite{liu2025multipath}, yielding the magnitude $\hat{v}_\mathrm{ma}^s$ and direction $\hat{\theta}_\mathrm{ma}^s$. 
The estimated velocity vector is then given by $\left(\hat{v}_\mathrm{ma}^s\cos{(\hat{\theta}_\mathrm{ma}^s)},\hat{v}_\mathrm{ma}^s\sin{(\hat{\theta}_\mathrm{ma}^s)}\right)$.

\textbf{Step 4:} Based on \textbf{\textcolor{blue}{Steps} 1-3}, we obtain raw position and velocity estimates over $S$ sensing snapshots. 
To obtain the acceleration and improve the smoothness of estimation, a six-dimensional Kalman filter with a CA model is applied to output physical information matrix $\mathbf{P}_\mathrm{tv}$, which serves as input to the subsequent BIP inference model.
Although real motion does not strictly follow CA, the CA model can provide an effective local linear approximation.

\begin{figure}
    \centering
    \includegraphics[width=0.35\textwidth]{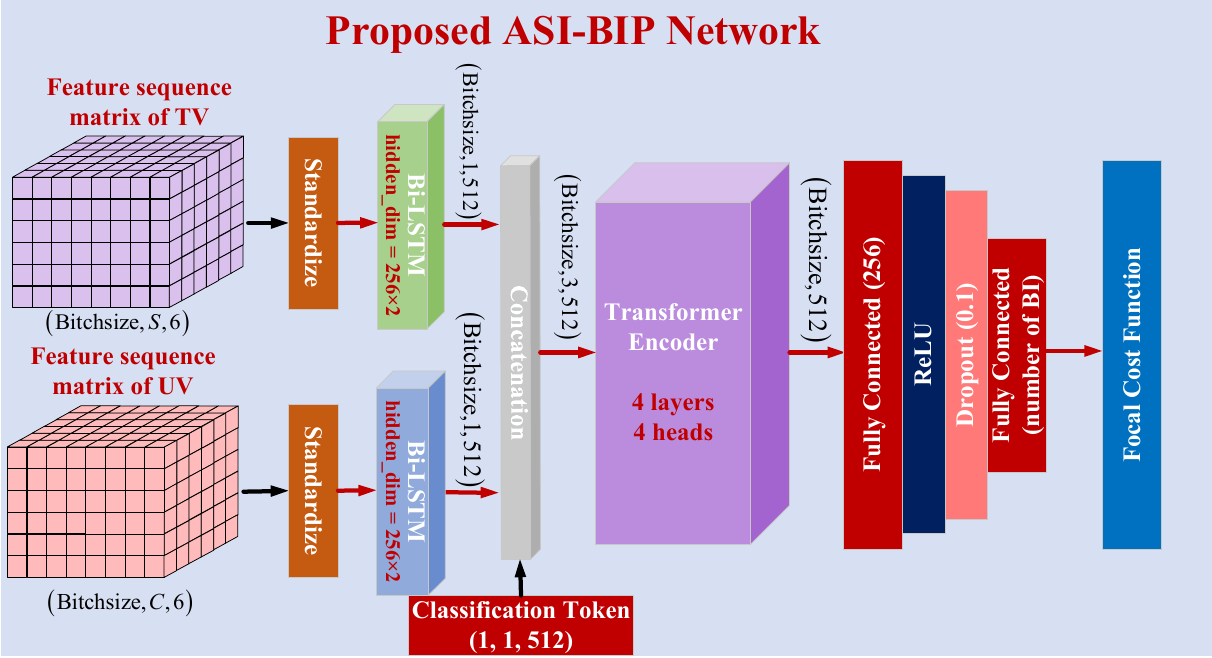}
    \caption{\underline{\textbf{(Modified)}} Proposed ASI-BIP network architecture.}
    \label{fig2}
\end{figure}

\subsection{Proposed ASI-BIP Network}
As illustrated in Fig.~\ref{fig2}, we propose the ASI-BIP network to model spatiotemporal interactions between asynchronous physical information matrices and enable accurate BIP. 



\textbf{1) Dual independent Bi-LSTMs:} Direct alignment or interpolation of matrices with different refresh rates may distort dynamics. We adopt a dual-pathway design: a high-rate branch for TV's motion and a low-rate branch for UV's behavior.
\textbf{2) Transformer encoder:} To capture inter-vehicle interaction, a Transformer encoder applies self-attention over the dual LSTM outputs~\cite{Fang2024BIP}, dynamically weighting cross-vehicle dependencies.
\textbf{3) Classification token:} We introduce a learned classification token to refine complex interaction information and capture global context for the final classifier.

For the cost function, we define the label as the ordinal index of each category in a set of $G$ behavioral intention classes.
Given \textcolor{blue}{that} the number of samples and classification difficulty of different categories are inconsistent, we adopt the focal cost function~\cite{Lin_2017_ICCV}
\begin{equation}\label{eq4}
{\fontsize{8}{8} L_{total}=\frac{1}{E}\sum_{e=1}^{E}\left[  -\sum_{g=1}^Gy_g\alpha_g\left(1-p_g\right)^\gamma \log\left(p_g\right)\right],}
\end{equation}
where $E$ denotes the number of samples and $g=\left\{1,2,\cdots,G\right\}$ denotes the index of intention classes; $y_g=1$ if $g$ is the true class, else 0; $p_g$ is the predicted probability; $\alpha_g$ and $\gamma$ are class-weighting and focusing parameters.

\section{Simulation Results and Analysis}\label{se4}
We evaluate the feasibility, superiority, and generalization ability of the proposed ISAC-based BIP framework.

\subsection{Simulation Setup}
We consider the 7 intentions defined in Section~\ref{se3-a}.
The ``\textit{following}'', which denotes a benign driving state, is intentionally excluded from the training set and subsequently used for generalization assessment as an unknown behavioral intention.

To build the dataset, we first generate 5-second physical information matrices for the TV and UV using a parameterized kinematic model~\cite{Fang2024BIP}. 
The refresh rate for ISAC cooperative sensing is 400 Hz, while the refresh rate for the UV's onboard sensors is 100 Hz~\cite{Benrachou2022,liu2025cooperative}. 
Based on data characteristics, the observation window is set to 2.2 s.
For each behavior, the kinematic parameters are randomly sampled within predefined ranges to reflect the uncertainty of the real-world. 
We then synthesize ISAC echo signals corresponding to the TV's physical information matrix with the following settings: Signal-to-noise ratio (SNR) $\in[0,10]$ dB, carrier frequency $f_\mathrm{c} = 3.5$ GHz, subcarrier spacing $\Delta f = 30$ kHz, $N_\mathrm{c}=128$ subcarriers, $M=42$ OFDM symbols, and $N_\mathrm{R} = 8$ antennas~\cite{Li2023,Liu2024target,liu2025multipath}.
\textcolor{blue}{$\mathbf{z}_{n,m}^{i,s}$ is modeled as additive white Gaussian noise.}
The final dataset comprises 6,000 samples, each labeled by the ordinal index of its behavioral category.

The dataset is split into 80\% training and 20\% validation sets, and a statistically co-distributed, mutually exclusive independent test set of 1,200 samples is used for final evaluation.
\textcolor{blue}{Models are trained with a batch size of 32, a learning rate of 0.0001},
and the Adam optimizer ($\beta_1 = 0.9$, $\beta_2 = 0.999$). 
All experiments are implemented in Python 3.12 with PyTorch 2.4.1 and CUDA 12.1 on an NVIDIA RTX 4090 GPU.

\subsection{Feasibility of the Proposed Framework}
Fig.~\ref{fig3} demonstrates the focal cost function and confusion matrix of the proposed framework, showing the feasibility of the proposed framework.
We adopt the F1-score and the average predicted accuracy as the performance \textcolor{blue}{metrics} for each behavioral intention $g$~\cite{Powers2011}
{\fontsize{8}{8}\begin{equation}\label{eq5}
    \text{F1-score}(g) = \frac{2P(g)R(g)}{P(g)+R(g)}, \quad  P_\mathrm{av}^g = \frac{\sum_{n=1}^{N_\mathrm{val}}\mathbb{I}\left(\hat{Q}_n^g=Q_n^g\right)}{N_\mathrm{val}}
\end{equation}}
\noindent
where precision $P(g)=\frac{T_\mathrm{P}(g)}{T_\mathrm{P}(g)+F_\mathrm{P}(g)}$ and recall $R(g) = \frac{T_\mathrm{P}(g)}{T_\mathrm{P}(g)+F_\mathrm{N}(g)}$. $T_\mathrm{P}(g)$, $F_\mathrm{P}(g)$, and $F_\mathrm{N}(g)$ denote the true positives, false positives, and false negatives, respectively~\cite{Powers2011}.
$N_\mathrm{val}$ is the number of samples in the validation set, $Q_n^g$ and $\hat{Q}_n^g$ are the true and predicted behavioral intentions for the $n$-th sample, respectively, and $\mathbb{I}(\cdot)$ is the indicator function.

Fig.~\ref{fig3.a} demonstrates effective model training, as both training and validation losses converge steadily without overfitting. 
Fig.~\ref{fig3.b} further corroborates the model's viability, showing \textcolor{blue}{that} the validation accuracy quickly rises and stabilizes near 90\%. 
In Fig.~\ref{fig3.c}, the F1-score for ``\textit{hard brake}'' (0.7568) is the lowest, while ``\textit{overtake}'' achieves a perfect score (1.0000). This disparity arises because ``\textit{hard brake}'' events are brief and less distinct, while ``\textit{overtake}'' events are prolonged and exhibit more distinguishable features.
In particular, ``\textit{hard accel}'' (0.9333) is much higher than ``\textit{hard brake}''. This is because ``\textit{hard accel}'' often occurs as a sustained, proactive maneuver, resulting in longer and more discernible feature sequences that are easier for the model to capture.

\begin{figure}
	\centering
    \subfigure[ Focal cost function] {\label{fig3.a}\includegraphics[width=0.22\textwidth]{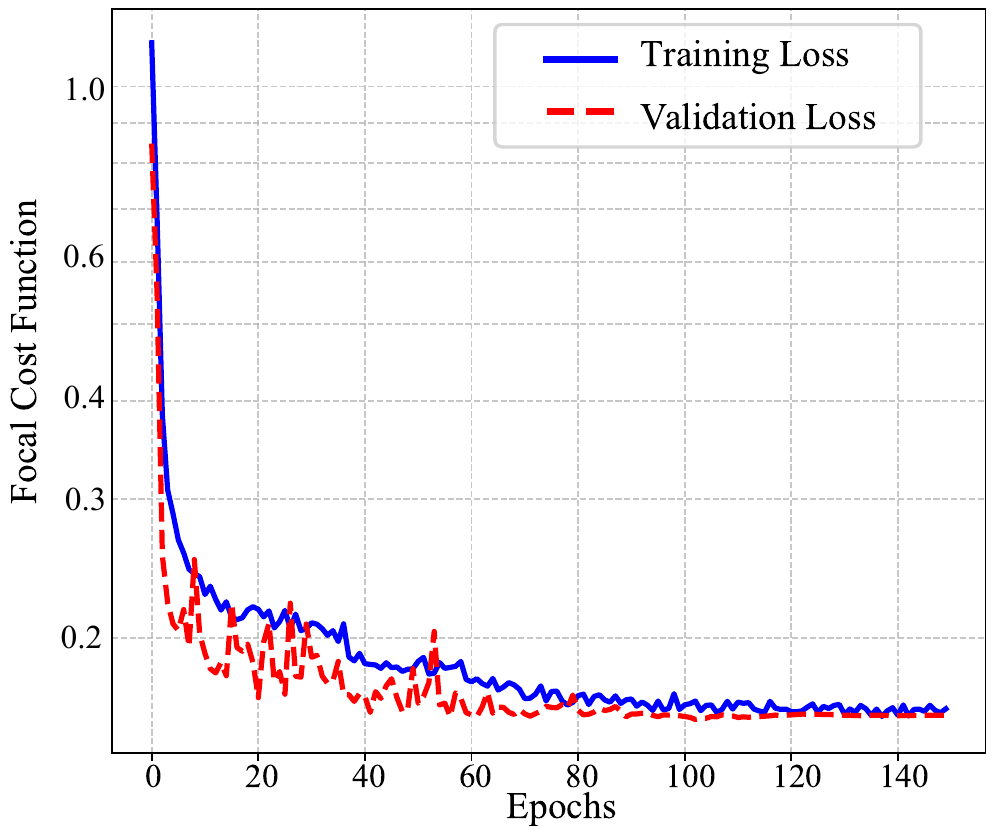}}
    \subfigure[ Average predicted accuracy]{\label{fig3.b}\includegraphics[width=0.22\textwidth]{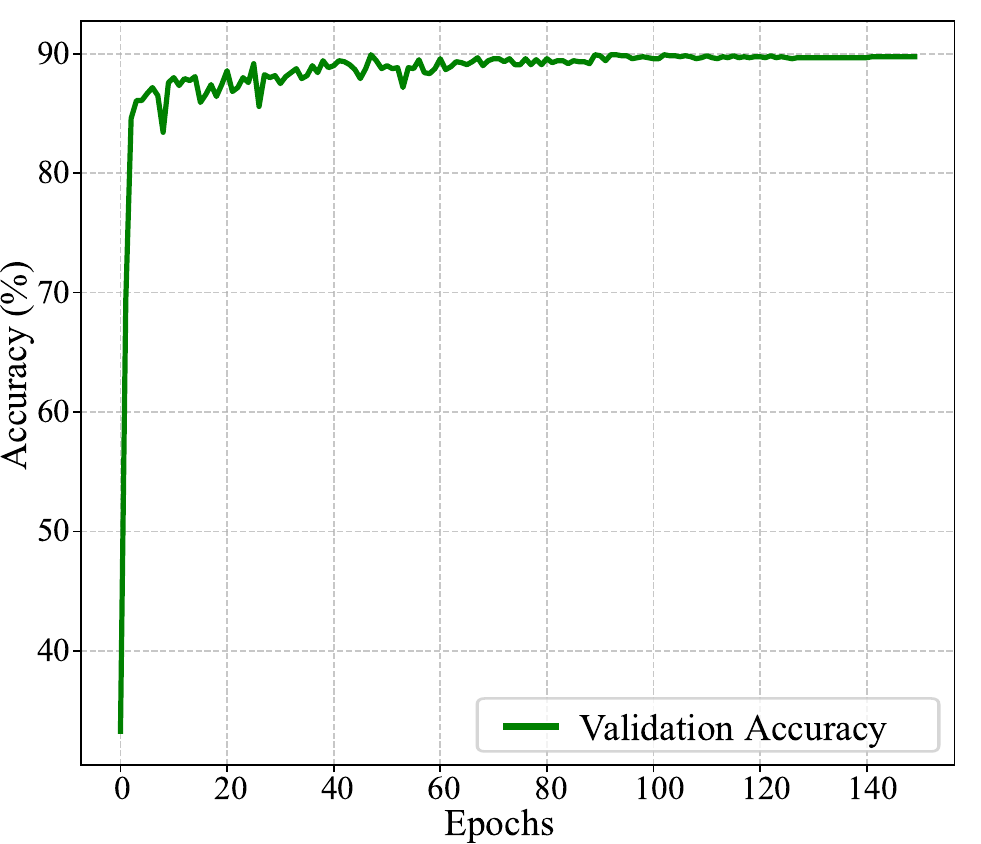}}
    \subfigure[Confusion matrix] {\label{fig3.c}\includegraphics[width=0.3\textwidth]{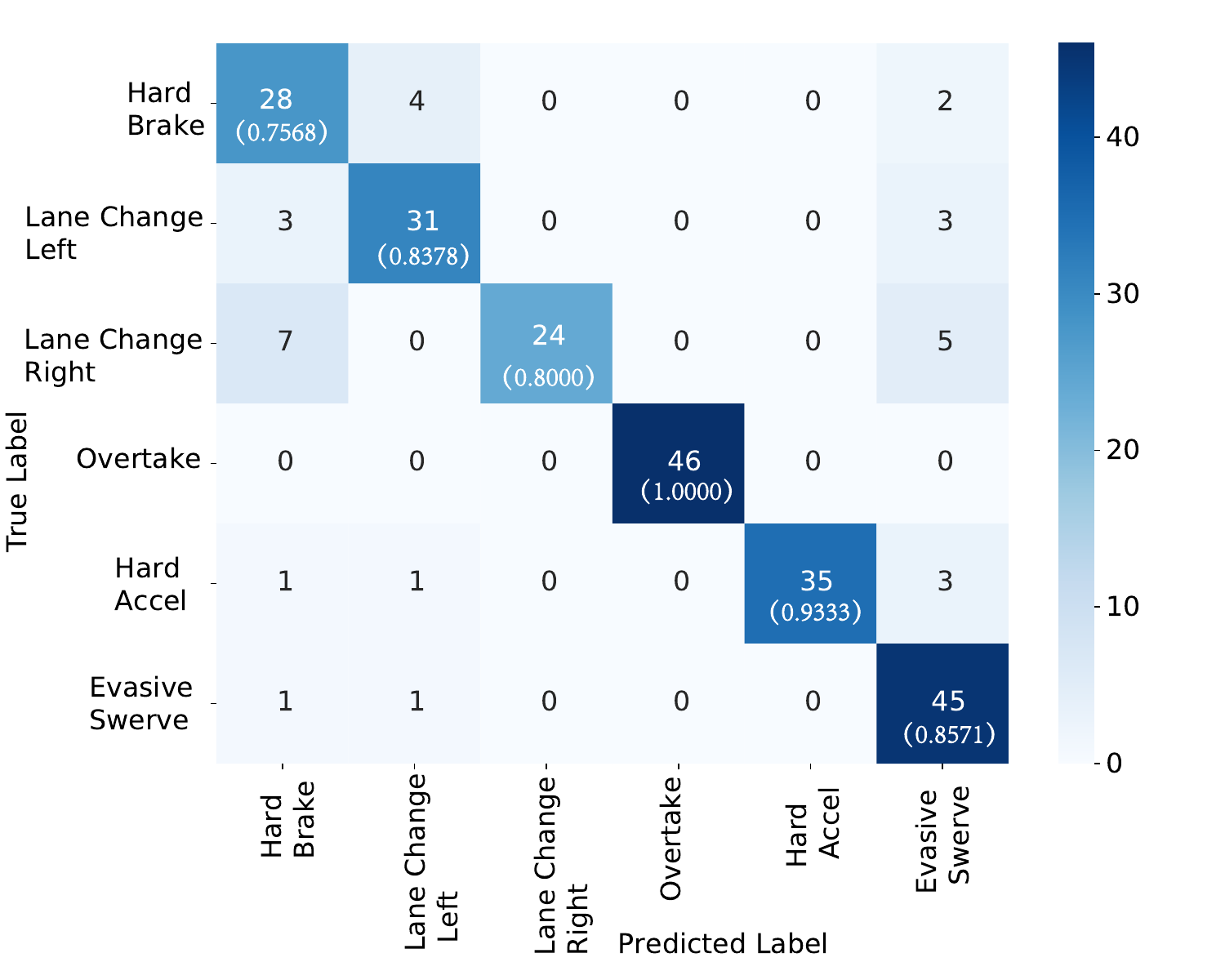}}
	\caption{ Loss and confusion matrix of the proposed framework.}
	\label{fig3}
\end{figure}

\subsection{Generalization ability of the Proposed Framework}
Fig.~\ref{fig4} assesses the model's ability to generalize to the unseen ``\textit{following}'' using the receiver operating characteristic (ROC) curve, prediction confidence distribution, and confusion matrix.
In Fig.~\ref{fig4.a}, the ROC curve for distinguishing ``\textit{following}'' from the six known intentions yields an area under the curve (AUC) of 0.6994, indicating a modest but non-trivial capability for novel intention detection.
The confidence distributions in Fig.~\ref{fig4.b} provide a visual explanation for this performance, showing that while known intentions generally receive higher confidence scores, a considerable overlap exists between the distributions for known and unknown intentions.
Finally, the confusion matrix in Fig.~\ref{fig4.c} (at a fixed decision threshold) reveals that most ``\textit{following}'' samples are correctly identified (last row), but certain known behaviors, particularly ``\textit{hard brake}'' and ``\textit{evasive swerve}'', are often misclassified as unknown (last column).
Collectively, these results demonstrate that the framework exhibits preliminary generalization to unseen intentions, though performance remains suboptimal and warrants further improvement.
\begin{figure}
	\centering
    \subfigure[ ROC curve for ``\textit{following}'' detection] {\label{fig4.a}\includegraphics[width=0.222\textwidth]{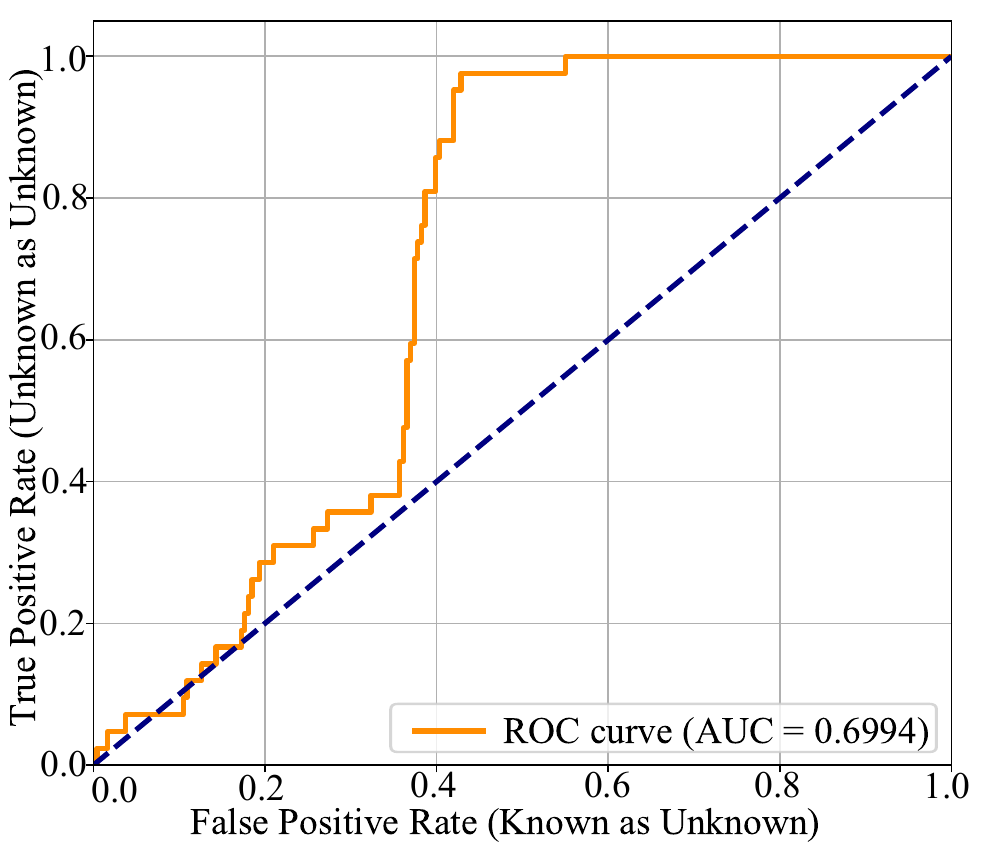}}
    \subfigure[ Distribution of prediction confidence]{\label{fig4.b}\includegraphics[width=0.22\textwidth]{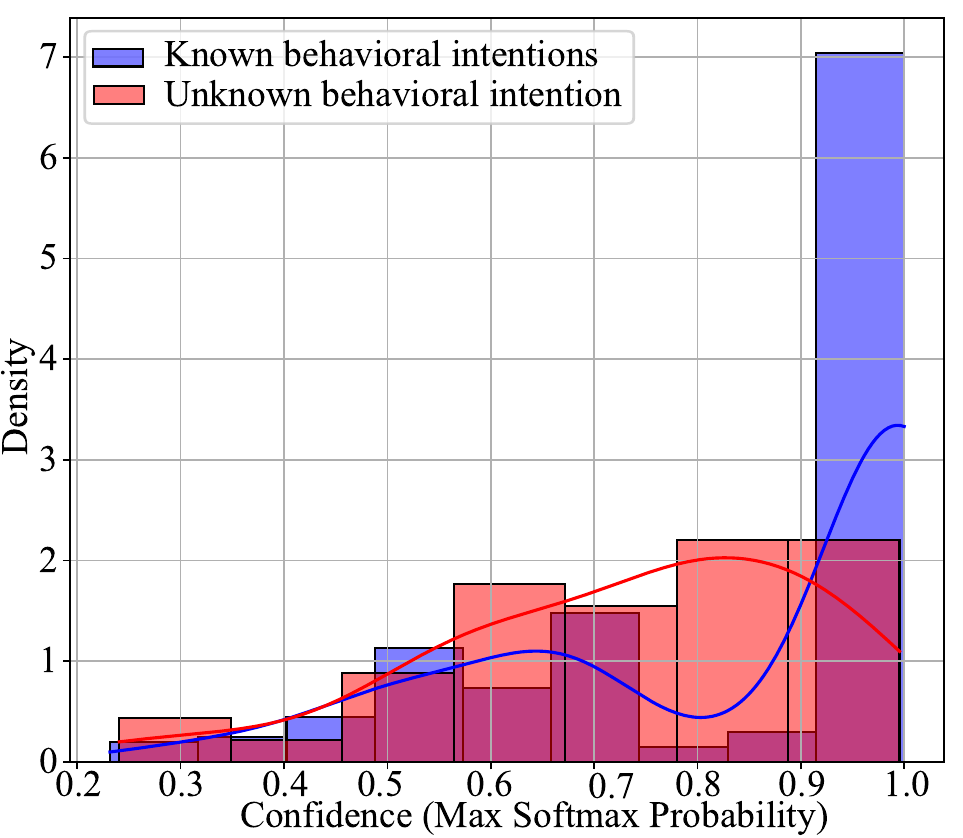}}
    \subfigure[Confusion matrix based on a threshold of 0.99] {\label{fig4.c}\includegraphics[width=0.3\textwidth]{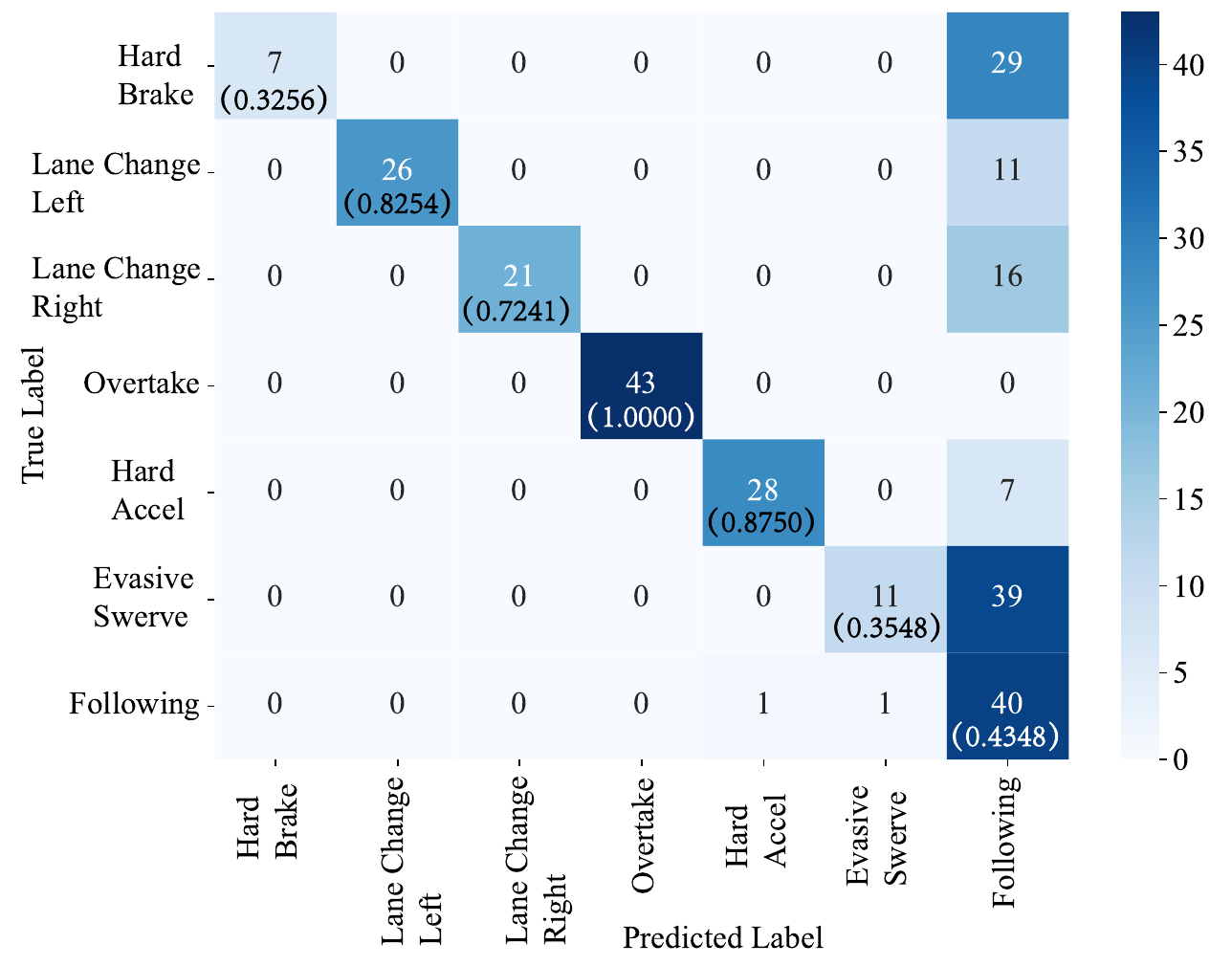}}
	\caption{ Generalization ability of the proposed framework for unknown behavioral intention}
	\label{fig4}
\end{figure}
Then, we assess the model's generalization capability against unseen SNR levels.
Since the model was trained using a dataset sampled only \textcolor{blue}{at} SNR $\in[0,10]$ dB, we evaluate its performance across a wider SNR range from $-10 \mathrm{~dB}$ to $20 \mathrm{~dB}$ using the same independent test set.
As shown in Fig.~\ref{fig5}, the F1-score for ``\textit{overtake}'' remains consistently at 1.0 across all SNR values. 
This is attributed to ``\textit{overtake}'' being a long-term, interactive behavioral intention, whose discriminative features come from both UV and TV, while the information input from UV is not directly affected by the SNR. 
Therefore, the ASI-BIP network is able to focus its attention on mining the interactive features of UV, thereby achieving immunity to noise. 
However, the occurrence of this phenomenon is predicated on the network having a sufficiently powerful feature extraction capability and the intention to overtake itself being clear enough.
In contrast, other behavioral intentions exhibit a common trend: their F1-scores improve with increasing SNR and eventually saturate. 
These results confirm that the proposed model maintains reliable prediction performance under unseen SNR conditions, demonstrating strong generalization capability.

\begin{figure}
    \centering
    \includegraphics[width=0.30\textwidth]{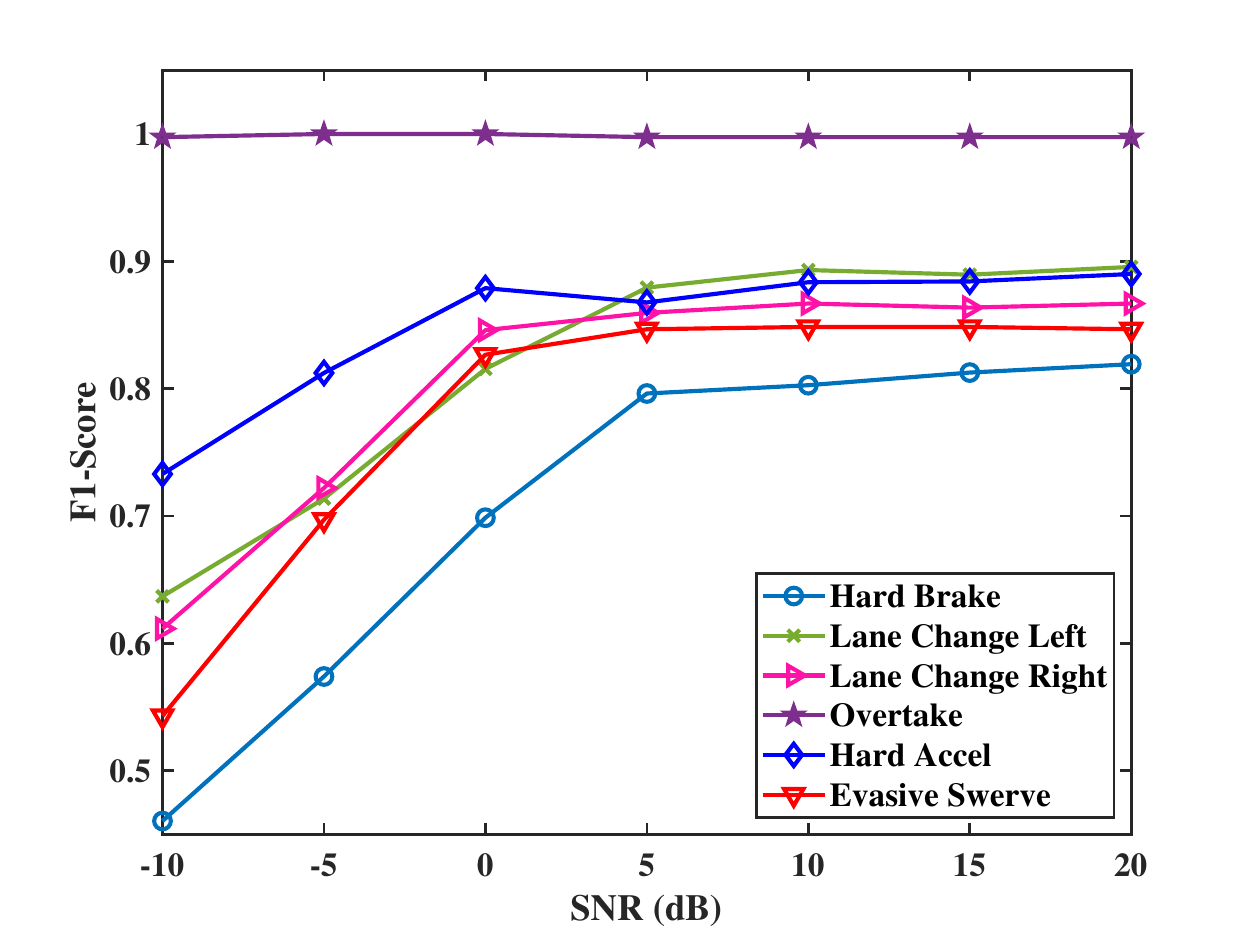}
    \caption{The F1-scores of various behavioral intentions with different SNR}
    \label{fig5}
\end{figure}

\subsection{Superiority of the Proposed Framework}
To evaluate the advantages of ISAC-based BIP over conventional onboard millimeter-wave radar in challenging environments, we consider five comparative cases in Table~\ref{tab1}: Case 1, a ground-truth baseline; Case 2, the proposed ISAC-based framework; Case 3, onboard millimeter-wave radar in a clear environment; Case 4, onboard millimeter-wave radar in a partial NLoS scenario, simulated by randomly dropping three-quarters of the trajectory data; and Case 5, onboard millimeter-wave radar in adverse weather (echo power loss of 20 dB).
To ensure a fair comparison, Case 2 is evaluated under the same adverse weather as Case 5. The NLoS scenario is excluded for ISAC, as its inherent multi-view advantage renders a direct comparison less meaningful.
It is noted that all performance metrics for the baseline cases (Case 1, 3, 4, and 5) are generated by re-implementing their respective sensing or inference modules within our uniform simulation framework, ensuring a direct comparison.

The analysis of Table~\ref{tab1} reveals several key insights. The 5.25\% accuracy gap between Case 1 and Case 2 indicates potential for enhancement in our ISAC echo signal processing stage. 
Furthermore, the performance of Case 2 is comparable to Case 4, with both being surpassed by Case 3. This is attributed to the higher intrinsic accuracy of onboard sensors under LoS conditions, combined with the Transformer's capability to mitigate NLoS effects through robust contextual reasoning. 
Critically, Case 2 significantly outperforms Case 5 with an 11.4\% improvement in the F1 score, an advantage attributed to the lower path attenuation of the sub-6 GHz band used by ISAC in adverse weather.

\begin{table}[h!]
    \centering
    \caption{Comparison of model performance under different cases, where the average accuracy and macro F1-score denote the average BIP accuracy rate and F1-score of each behavioral intention, respectively.}
    \label{tab1}
    \resizebox{0.25\textwidth}{!}{%
        \begin{tabular}{l c c}
            \toprule
            \textbf{Cases} & \textbf{Average accuracy} & \textbf{Macro F1-score} \\
            \midrule

            Case 1       & 92.33\% & 0.9249 \\
            \textbf{Case 2}         & 87.08\% & 0.8642 \\
            Case 3      & 87.50\% & 0.8761 \\
            Case 4        & 86.42\% & 0.8660 \\
            Case 5     & 79.17\% & 0.7758 \\ 
            \bottomrule
        \end{tabular}
    }
\end{table}

\section{Conclusion}\label{se5}
This letter introduces an ISAC-based BIP framework, leveraging the robust sensing capabilities of ISAC to achieve event-level sensing. 
We propose the ASI-BIP network, which uniquely utilizes dual independent Bi-LSTMs and a Transformer to realize feature extraction from asynchronous data and effectively capture spatiotemporal vehicle interactions. Extensive simulations demonstrate that our framework significantly outperforms conventional onboard sensors, achieving an 11.4\% improvement in the macro F1-score under adverse weather conditions. This work not only confirms the feasibility and superiority of ISAC for event-level sensing but also highlights its potential to enable robust intelligent systems in applications \textcolor{blue}{such as} smart cities, factories, and transportation.

\bibliographystyle{IEEEtran}
\bibliography{reference}

\end{document}